\documentstyle[aps,prb]{revtex}

\def\sqr2{\sqrt{2}}
\def\half{\frac{1}{2}}

\def\>{\rangle}
\def\<{\langle}
\def\X{X}
\def\Y{Y}
\def\L{{\cal L}}

\def\dotPsi{\dot{\Psi}}
\def\dotnu{\dot{\nu}}
\def\ddotPsi{\ddot{\Psi}}
\def\ddotnu{\ddot{\nu}}
\def\ddotlambda{\ddot{\lambda}}

\def\RPAl{\langle {\rm RPA} |}
\def\RPAr{|{\rm RPA} \rangle}

\begin{document}
%\draft
\tightenlines
%\twocolumn[\hsize\textwidth\columnwidth\hsize\csname@twocolumnfalse\endcsname

%%%%%%%%%%%%%%%%%%%%%%%%%%%%%%%%%%%%%%%%%%%%%%%%%%%%%%%%%%%%%%%%%%%%%%%%%%%%%
\title{Car-Parrinello molecular dynamics on excited state surfaces }
%%%%%%%%%%%%%%%%%%%%%%%%%%%%%%%%%%%%%%%%%%%%%%%%%%%%%%%%%%%%%%%%%%%%%%%%%%%%%

%%%%%%%%%%%%%%%%%%%%%%%%%%%%%%%%%%%%%%%%%%%%%%%%%%%%%%%%%%%%%%%%%%%%%%%%%%%%%
\author{Eric R. Bittner and D.S. Kosov \footnote{
Present address: Department of Chemistry, UMIST, 
Manchester, M60 1QD, UK}} 
%%%%%%%%%%%%%%%%%%%%%%%%%%%%%%%%%%%%%%%%%%%%%%%%%%%%%%%%%%%%%%%%%%%%%%%%%%%%%

%%%%%%%%%%%%%%%%%%%%%%%%%%%%%%%%%%%%%%%%%%%%%%%%%%%%%%%%%%%%%%%%%%%%%%%%%%%%%
\address{Department of Chemistry \\
University of Houston \\
Houston, TX 77204, USA} 
%%%%%%%%%%%%%%%%%%%%%%%%%%%%%%%%%%%%%%%%%%%%%%%%%%%%%%%%%%%%%%%%%%%%%%%%%%%%%

\date{ \today}
\maketitle
\begin{abstract}
This paper describes a method to do {\em ab initio} molecular dynamics
in electronically excited systems within the random phase
approximation (RPA).  Using a dynamical variational treatment of the
RPA frequency, which corresponds to the electronic excitation energy
of the system, we derive coupled equations of motion for the RPA
amplitudes, the single particle orbitals, and the nuclear coordinates.
These equations scale linearly with basis size and can be implemented
with only a single holonomic constraint.  Test calculations on a model
two level system give exact agreement with analytical results.
Furthermore, we examined the computational efficiency of the method by
modeling the excited state dynamics of a one-dimensional polyene
lattice.  Our results indicate that the present method offers a
considerable decrease in computational effort over a straight-forward
configuration interaction (singles) plus gradient calculation
performed at each nuclear configuration.
\end{abstract}
\maketitle

{\bf Pacs}: 71.15-m, 71.15.Pd, 31.50.+w
%\newpage
%\ ]

%\narrowtext
%%%%%%%%%%%%%%%%%%%%%%%%%%%%%%%%%%%%%%%%%%%%%%%%%%%%%%%%%%%%%%%%%%%%%%%%%%
\section{Introduction}
%%%%%%%%%%%%%%%%%%%%%%%%%%%%%%%%%%%%%%%%%%%%%%%%%%%%%%%%%%%%%%%%%%%%%%%%%%

In recent years, there has been considerable advances towards fully
{\em ab initio} molecular dynamics simulations using inter- and
intramolecular potentials derived from quantum many-body interactions.
This activity has been largely spurred by the introduction of hybrid
quantum/classical dynamics treatments, such as the density functional
theory/molecular dynamics methods introduced by Car and
Parrinello~\cite{MParrinello85,MParrinello89,MParrinello91} (CPMD).
In this approach, the quantum electronic orbitals, $\{\phi_i(1)\}$,
are coupled to the classical nuclear degrees of freedom, ${\bf R}_n$,
through a fictitious Lagrangian, which along with an appropriate set of
Lagrange multipliers, $(\Lambda_{ij})$, leads to a set of equations of
motion
\begin{eqnarray}
\mu \ddot{\phi}_i(1) &=& -\frac{\delta E[\{\phi_i\},{\bf R}_n]}{\delta
\phi_i^*(1)} + \sum_k \Lambda_{ik}\phi_k(1), \nonumber \\
M_n\ddot{\bf R}_n &=& -\nabla_{\bf R}E[\{\phi_i\},{\bf R}_n],\label{eq:CP}
\end{eqnarray}
where $ E[\{\phi_i\},{\bf R}_n]$ is the Hohenberg and Kohn energy
functional~\cite{Kohn64}.  This approach has proven to be highly
successful in simulating chemical dynamics in the condensed phase.

However, there are a number of distinct disadvantages to the method.
First, the equations of motion involve both fast (the orbitals) and
slow (the nuclei) degrees of freedom. Thus, the computational
time-step must be rather small.  Secondly, imposing the holonomic
constraint of orbital orthogonality forces a single iteration of the
approach to scale as $\sim M \times N^2$ where $N$ is the number of
particles and $M$ the size of the electronic basis.  Finally, the
method can only treat systems in the electronic ground state.

Several groups have attempted to overcome each of these difficulties.
The time-step problem has been attacked via multiple time-scale
methods~\cite{TP94a,TP94b} and by using the electronic density
instead of the Kohn-Sham~\cite{Kohn65} orbitals as the dynamical
variable~\cite{Pearson}.  Several algorithms have been introduced to
tackle the scaling problem~\cite{MParrinello91}. In these cases,
scaling more favorable than $\sim N^3$ is achieved only at the cost
of several assumptions and approximations.

Recently, Alavi, Kohanoff, Parrinello, and Frenkel have developed a
density functional based method for treating systems with finite
temperature electrons~\cite{MParrinello94,Par96}.  This approach has a
number of attractive features, foremost being the inclusion of
fractional occupation of the electronic orbitals.  Dynamics in this
scheme are isothermal rather than adiabatic, allowing for incoherent
transitions between electronically excited states.  While the approach
works well for small bandgap systems (such as metals or  dense
hydrogen), this approach is not useful for systems with larger band
gaps ($\Delta E \gg kT$) or photo-excited systems.

In this paper, we present an alternative formulation of the
Car-Parrinello method for treating electronically excited systems.  We
begin by introducing the general computational algorithm for the
dynamical optimization of excited electronic states in the presence of
``classically propagating'' ions. This general theoretical scheme is
developed without specific reference to the description of the
electronic excited state.  The principal results of this section are a
set of dynamical equations of motion for the electronic amplitudes and
nuclear coordinates in a generalized phase space.  We then derive
specific equations of motion for the electronic amplitudes and nuclear
coordinates under the Random Phase Approximation (RPA) for the
electrons.  Finally, we
demonstrate the salient features of the approach through a simplified
SU(2) model of interacting electrons and via a realistic model of conjugated 
one-dimensional polymer lattices.

%%%%%%%%%%%%%%%%%%%%%%%%%%%%%%%%%%%%%%%%%%%%%%%%%%%%%%%%%%%%%%%%%%%%%%%
\section{Classical equations of motion for the dynamical optimization
 of  excited quantum states}
%%%%%%%%%%%%%%%%%%%%%%%%%%%%%%%%%%%%%%%%%%%%%%%%%%%%%%%%%%%%%%%%%%%%%%%

We begin by developing a general dynamical theory for molecular
dynamics in excited electronic states.  We first define the
$N$-electron Fock space as being parameterized by a set of dynamical
variables $\lambda(\tau)$ which include the nuclear positions and any
other dynamical variables and ensemble constraints placed upon the
system (e.g. pressure, temperature, volume etc...):
\begin{eqnarray}
{\cal F} =
\left\{ |n_1,n_2, ...(\lambda)\rangle, \sum_{i} n_i =N\right\}
\end{eqnarray}
We  describe the ground state wave function as an expansion on
the basis in ${\cal F}$:
\begin{eqnarray}
|\Psi_o (\lambda)\> =\sum_{1...i...} C_{i}(\lambda)
 |n_1 ... n_i... (\lambda)\>
\; ,
\<\Psi_o(\lambda)|\Psi_o(\lambda) \>=1
\end{eqnarray}
$|\Psi_o(\lambda) \>$ is not a Slater determinant in general case.

We next define the mapping operators, $P^\dagger_\nu$, which act on the
ground state to produce a new state, $|\nu (\lambda)\>$, which is   orthogonal to 
 $|\Psi_o(\lambda) \>$  in the subspace ${\cal F}_+$ of the Fock space:
\begin{eqnarray}
|\nu(\lambda) \> =P^{\dagger}_{\nu}|\Psi_o (\lambda)\>,
\end{eqnarray}
\begin{eqnarray}
P_{\nu} |\Psi_o (\lambda)\>
=0,
\end{eqnarray}
\begin{eqnarray}
\<\nu(\lambda)|\Psi_o(\lambda)\> =
 \<\Psi_o(\lambda)|P_\nu |\Psi_o(\lambda)\> = 0 \;, 
\<\nu(\lambda)|\nu(\lambda) \> =1
\end{eqnarray} 
If the $|\Psi_o(\lambda) \>$ is 
a true  ground state of the $N$-particle system, 
the excited states should lie in the orthogonal subspace ${\cal F}_+$
\cite{Messiah}.

Variational determinations of the excited states are thus searches
in the orthogonal subspace for the optimal mapping operator
$P^{\dagger}_{\nu}$ and for the optimal basis of the Fock space 
$|n_1 n_2 ...n_i .. \> $
such that the energy functional
\begin{eqnarray}
\varepsilon_{\nu}&=& \<\nu(\lambda)|H|\nu(\lambda)\>  ,
\nonumber 
\\
&=&
\<\Psi_o(\lambda)|H|\Psi_o(\lambda)\>
\nonumber
\\ &+&
\left( \<\nu(\lambda)|H|\nu(\lambda)\>
-\<\Psi_o(\lambda)|H|\Psi_o(\lambda)\> \right)  ,
\nonumber
\\
&=&\<\Psi_o(\lambda)|H|\Psi_o(\lambda)\> 
+\<\Psi_o(\lambda)|\left[ P_{\nu},\left[H,P^{\dagger}_{\nu}\right] \right]
|\Psi_o(\lambda)\>,
\nonumber
\\
&=& E_o +\omega_{\nu},
\end{eqnarray}
is at a variational minimum. We write the excitation energy, 
$\varepsilon_{\nu}$, as a functional of  both the ground 
and excited state
and introduce explicitly via the double commutator term the
energy gap,
$\omega_{\nu}$, between the ground and the excited state.

According to the calculus of variations, we can derive equations 
of motion which dynamically optimize the
excited state system.  We first
define formally a set of ``velocities'' of the states in the Fock space
\begin{eqnarray}
{|\dotnu\>} =\frac{d}{d\tau}|\nu\>
 \; , \;  {|\dotPsi_o\>} =\frac{d}{d\tau}|\Psi_o\>,
\end{eqnarray}
where $\tau$ is not the true physical time but rather a parameter
which characterizes the evolution of the amplitudes through the
Hilbert space subject to the orthogonalization constraints given
above.  The ``classical'' Lagrangian for our system is given by
\begin{eqnarray}
\L&=&\frac{1}{2} \mu {|\dotPsi_o\>}{ \<\dotPsi_o|}
+\frac{1}{2} \mu'
{|\dotnu\>}{ \<\dotnu|}
+\frac{1}{2}m\dot{\lambda}^2(\tau)\nonumber
\\
&-&\varepsilon_{\nu}[|0\>,|\nu\>,\lambda]  \nonumber \\
&-&\Lambda (\<\Psi_o|\Psi_o\>-1) -\Gamma (\<\nu|\nu\>-1),
\end{eqnarray}
where $\Lambda$ and $\Gamma$ are Lagrange multipliers
arising from the holonomic ortho-normalization
constraint for the ground and excited states and which 
allow $|\nu\>$ and $|\Psi_o\>$ to be treated as independent variables.
The masses $\mu$ and $\mu'$ are fictitious masses and have no true
physical meaning.  This Lagrangian leads to the classical
Euler-Lagrange equations~\cite{Goldstein} for the wave functions:
\begin{eqnarray}
\frac{d}{d\tau} \frac{\partial \L}{\partial{|\dotPsi_o\>}}-
 \frac{\partial \L}{\partial |\Psi_o\>} =0,
\\
\frac{d}{d\tau} \frac{\partial \L}{\partial{|\dotnu\>}}-
 \frac{\partial \L}{\partial |\nu\>}=0.
\end{eqnarray}
The Euler-Lagrange equation dictates the following 
equations of motion for quantum dynamical optimization in the Hilbert space:
\begin{eqnarray}
\mu{|\ddotPsi_o\>}&=&-\frac{\delta\varepsilon_{\nu}}{\delta \<\Psi_o|} +
\Lambda |\Psi_o \> ,  \nonumber \\ 
\mu'{|\ddotnu\>}&=&-\frac{\delta \varepsilon_{\nu}}{\delta \<\nu|} +
\Gamma |\nu \>,  \nonumber \\
m{\ddotlambda} &=& -\frac{\delta \varepsilon_{\nu}}{\delta \lambda}.
\end{eqnarray}
This dynamical view is the
starting point for the remainder of our calculations.

%%%%%%%%%%%%%%%%%%%%%%%%%%%%%%%%%%%%%%%%%%%%%%%%%%%%%%%%%%%%%%%%%%%%
\section{Random Phase Approximation for Electronic Motions}
%%%%%%%%%%%%%%%%%%%%%%%%%%%%%%%%%%%%%%%%%%%%%%%%%%%%%%%%%%%%%%%%%%%

For a molecular system with $N$ electrons and $n$ nuclei, 
the full $N$-body Hamiltonian for a fixed set of nuclear positions 
$\{{\bf R}_\nu\}$, is given as (atomic units are used throughout)
\begin{eqnarray}
H&=&\frac{1}{2} \sum_{i} {\nabla_{i}}^{2}
-\sum_{\nu} \sum_{i} \frac{Z_{\nu}}{|{\bf R}_{\nu}-{\bf r}_{i}|}
+\sum_{ij}  \frac{1}{|{\bf r}_{i}-{\bf r}_{i}|}, \\
&=& \sum_i^N h_\circ^{(i)} + \sum_{ij} \frac{1}{|{\bf r}_{i}-{\bf r}_{i}|},
\end{eqnarray}
where $h_\circ^{(i)}$ is 
the one-body electron operator which includes
the electron kinetic energy and
the interaction between the electron and the
nuclear cores.
We define the  single-particle Hamiltonian, $H_1$, 
through either the Hartree-Fock (HF) approximation~\cite{Negele}
or via the density functional theory of Kohn and Sham~\cite{Kohn65}.
In either case, $H_1$ is a functional of the single particle density 
\begin{eqnarray}
\rho(12)=\sum_i^{occ}\phi_i^*(1)\phi_i(2).
\end{eqnarray}
The single particle states, $\{\phi_i(1)\}$, are  solutions
of a nonlinear
Schr\"odinger equation, 
\begin{eqnarray}
H_1[\rho,{\bf R}] |\phi_i\> = E_i[{\bf R}] |\phi_i\>,
\end{eqnarray}
which must be determined  at the start of the
calculation. 
Here $E_i[{\bf R}]$ is the orbital energy. The HF spin-orbital
wave functions and energies differ from the KS ones in the way the
terms in the single-particle Hamiltonian are treated.
 We will denote the Slater determinant wave-function
constructed from occupied single-particle HF (of KS) orbitals as
$|HF\>$ and note that this state represents either the Hartree-Fock
ground state or the Kohn-Sham ground state.

Having obtained a basis of single-particle states, we can write the
two-body Hamiltonian in second quantized form:
\begin{eqnarray}
H&=&
E_{HF}+\sum_{i
\sigma}
E_{i} :
a^{\dagger}_{i
\sigma}
a_{i
\sigma}:
\nonumber \\
&+& \frac{1}{2} \sum_{ijkl} \sum_{\sigma \sigma'}
(ij|v|kl) 
: a^{\dagger}_{i \sigma} a^{\dagger}_{j \sigma'} 
  a_{l \sigma} a_{k \sigma'} : ,
\end{eqnarray}
where $E_{i }$ is the HF single-particle eigenenergy,
\begin{eqnarray}
(ij|v|kl) = \int d1 d2 \phi_i^*(1)\phi_j^*(2)v(12)\phi_k(1)\phi_l(2),
\end{eqnarray}
is the matrix element of Coulomb electron-electron 
interactions,
$:\cdots:$ denotes a normal ordered product
and $E_{HF}$ is the HF ground state energy. The operators
$a^{\dagger}_{i \sigma}$  ($a_{i \sigma}$) create
(annihilate) single electrons in the $\{i \sigma\}$ HF spin-orbital state,
where
$\sigma= \pm \frac{1}{2}$ is the spin index of the electrons.

It is convenient at this point to move to a particle/hole
representation and introduce the particle/hole quasi-particle operator,
\begin{eqnarray}
\alpha_{i\sigma}|HF\> = 0,
\end{eqnarray}
through its action on the HF vacuum, where $i=p,h$ represents
particles (occupied states above the Fermi energy)  and
holes (vacancies below the Fermi energy). 
These quasi-particles operators are related to the fermion
operators written above via
\begin{eqnarray}
a^{\dagger}_{i \sigma}= \theta(E_{i} -E_{f}) \alpha^{\dagger}_{i \sigma}
+\theta(E_{f}-E_{i}) \alpha_{i \sigma},
\end{eqnarray}
\begin{eqnarray}
a_{i \sigma}= \theta(E_{i} -E_{f}) \alpha_{i \sigma}
+\theta(E_{f}-E_{i}) \alpha^{\dagger}_{i \sigma},
\end{eqnarray}
where $\theta(x)$ is the Heaviside step function and $E_f$ is the
Fermi energy. 

We can expand $H$ in terms of the particle/hole operators and
neglect terms with an odd number of particle or hole terms (i.e. 
anharmonic interaction 
terms of the form $\sim \alpha^{\dagger} \alpha \alpha \alpha$
and $\sim \alpha^{\dagger} \alpha^{\dagger} \alpha^{\dagger} \alpha $),
\begin{eqnarray}
H&=& E_{HF}+\sum_{p \sigma} E_p \alpha^{\dagger}_{p \sigma} \alpha_{p \sigma}+ 
 \sum_{h \sigma} E_h \alpha_{h \sigma} \alpha^{\dagger}_{h \sigma}
\nonumber
 \\
&+&\frac{1}{2} \sum_{p_1 p_2 h_3 h_4}
(p_1 p_2|v|
 h_3 h_4)
 \nonumber
 \\
&\times&
\sum_{\sigma \sigma'}
\left( \alpha^{\dagger}_{p1 \sigma}
\alpha^{\dagger}_{p2 \sigma'}\alpha^{\dagger}_{h3 \sigma}\alpha^{\dagger}_{h4 \sigma'}
+h.c. \right) \nonumber \\
&+&
\sum_{p_1 h_2 h_3 p_4}
(p_1 h_2|v| h_3 p_4) 
\sum_{\sigma \sigma'}
 \alpha^{\dagger}_{p1 \sigma}
\alpha^{\dagger}_{h3 \sigma} \alpha_{h2 \sigma'} \alpha_{p4 \sigma'} 
\nonumber \\
&-&
\sum_{p_1 h_2 p_3 h_4}
(p_1 h_2|v| p_3 h_4) 
\sum_{\sigma \sigma'}
\alpha^{\dagger}_{p1 \sigma}
\alpha^{\dagger}_{h4 \sigma'}\alpha_{h2 \sigma'}\alpha_{p3 \sigma}
 \nonumber \\
&+& \mbox{\rm anharmonic terms},
\end{eqnarray}
where $p$ and $h$ refer to particle and hole single particle states.

Note that each ``harmonic'' term in our expansion of $H$ involves  
an even number of particle and hole operators.  We can form a new set
of exciton operators by taking 
bilinear combinations of the particle/hole 
operators with total angular momentum
$S$ and projection $m_S$,
\begin{eqnarray}
A^{\dagger}_{ph}(S m_S) =\sum_{\sigma_1 \sigma_2} 
\<\frac{1}{2} \sigma_{1} \frac{1}{2} \sigma_{2}|Sm_s \> 
\alpha^{\dagger}_{p \sigma_{1}} \alpha^{\dagger}_{ h
\overline{\sigma}_{2}} ,\nonumber \\
A_{ph}(S m_S) =\sum_{\sigma_1 \sigma_2} 
\<\frac{1}{2} \sigma_{1} \frac{1}{2} \sigma_{2}|S m_S\> 
 \alpha_{h \overline{\sigma}_{2}} \alpha_{h \sigma_1},
\end{eqnarray}
where $\<s_1 \sigma_1 s_2 \sigma_2 | Sm_s\>$ is a Clebsch-Gordon
coefficient. 
Our notation  is such that $\overline{\sigma}$ 
denotes the action of time reversal
operator on the hole state,
\begin{eqnarray}
 \alpha^{\dagger}_{ h \overline{\sigma}} =(-)^{\frac{1}{2}+\sigma} 
\alpha^{\dagger}_{ h  -\sigma} \; .
\end{eqnarray} 
The  $\{A^{\dagger}_{ph}(S m_S),A_{ph}(S m_S)\}$ serve to create (or
destroy) particle/hole exciton with spin $S$ and projection $m_s$. 

Writing just the  ``harmonic'' part of  $H$ in terms of exciton 
operators, 
\begin{eqnarray}
H_2&=& \sum_{p \sigma} E_p \alpha^{\dagger}_{p \sigma} \alpha_{p \sigma}+ 
 \sum_{h \sigma} E_h \alpha_{h \sigma} \alpha^{\dagger}_{h \sigma}
\nonumber
\\
&-&\frac{1}{2} \sum_{p_1 p_2 h_3 h_4}
(p_1 p_2|v|h_3 h_4)
\sum_{S m_s} (-)^{S} \left( A^{\dagger}_{p_1 h_4}(S m_s)
A^{\dagger}_{p_2 h_3}(S \overline{m_s})+ h.c. \right)
\nonumber
\\
&+&\frac{1}{2} \sum_{p_1 h_2 h_3 p_4}
(p_1 h_2|v|h_3 p_4) 
\sum_{S m_s}
\sum_{S^{'} m_s^{'}}
\left((-)^{S} +1 \right)
\left((-)^{S^{'}}+1 \right)
 A^{\dagger}_{p_1 h_3}(S m_s)
A_{p_4 h_2}(S^{'} m_s^{'}) 
\nonumber
\\
&-& \sum_{p_1 h_2 p_3 h_4}
(p_1 h_2|v|p_3 h_4)
\sum_{S m_s}
A^{\dagger}_{p_1 h_4}(S m_s)
A_{p_3 h_2}(S m_s), 
\end{eqnarray}
where 
$
A_{p h}(S \overline{m_s})=
(-)^{S+m_s} A_{p h}(S -m)
$ 
is  the time reversal exciton annihilation operator.
This defines our ``quasi-harmonic'' exciton Hamiltonian.

The exact commutation relation between the exciton operators is 
\begin{eqnarray}
&[&A_{ph}(S \mu), A^{\dagger}_{p'h'}(S' \mu')  ] 
\nonumber
\\
&=&
\sum_{\sigma_1 \sigma_2} \sum_{{\sigma_1}' {\sigma_2}'}
\< \frac{1}{2} \sigma_{1} \frac{1}{2} \sigma_2 | \lambda \mu \>
\< \frac{1}{2} \sigma_{1}'\frac{1}{2} \sigma_2' | \lambda' \mu'
\>\nonumber
\\ &\times &
( \alpha_{h \overline{\sigma_{2}}} 
\alpha^{\dagger}_{h' \overline{\sigma_{2}'}}
\delta_{pp'} \delta_{\sigma_1 {\sigma_1}'}-
\alpha^{\dagger}_{p' \sigma_{1}'}
\alpha_{p \sigma_{2} } 
\delta_{hh'} \delta_{\sigma_2 {\sigma_2}'} ).
\label{ex-comm}
\end{eqnarray} 
We apply the Wick theorem to the right hand side of Eq. \ref{ex-comm}
and then neglect the normal ordering contribution to this relation. This is
the RPA \cite{Rowe} and in this approximation the commutation relation
between the exciton operators is bosonic
\begin{eqnarray}
[A_{ph}(S \mu), A^{\dagger}_{p'h'}(S' \mu')  ]
\approx \delta_{S S'} \delta_{\mu \mu'} \delta_{h h'} \delta_{p p'}.
\end{eqnarray} 
We then perform a canonical Bogolubov transformation
\cite{Blaizot}
 for both singlet
($S=0$) and triplet ($S=1$) exciton bosons
\begin{eqnarray}
Q^{\dagger}_{i}(00) &=& \sum_{ph} \X^{i}_{ph} A^\dagger_{ph}(00) 
- \Y^{i}_{ph}A_{ph} (00) , \nonumber \\
Q^{\dagger}_{i}(1m_s) &=& \sum_{ph} \X^{i}_{ph} A^\dagger_{ph}(1 m_s) 
- \Y^{i}_{ph}A_{ph} (1 \overline{ m_s}) ,
\end{eqnarray}
where $\X^i_{ph}$ and $\Y^i_{ph}$ are the RPA  amplitudes which
are yet to be determined.  These operators create and destroy
harmonic excitations in the electronic degrees of freedom.
Physically, the singlet RPA excitations correspond to ``breathing'' modes of
the electron cloud and triple RPA excitations correspond to dipolar
oscillations of the electronic system about the nuclear frame.  In
what follows, we will focus our attention upon the triplet excitations, as
these are the states created upon photo-excitation of a singlet ground
state. Equations of motion for singlet RPA excitations can be obtained as
well.

The commutation  relations for the RPA excitation creation operators
goes as follows:
\begin{eqnarray}
[Q_{\mu i},Q_{\mu' i'}^\dagger] &=&
\sum_{ph}{\X^{i}_{ph}}^{*} \X^{i'}_{ph}
-{\Y^{i}_{ph}}^{*} \Y^{i'}_{ph} \nonumber 
\\&=& \delta_{i i'} \delta_{\mu  \mu'}.
\label{RPA-orth}
\end{eqnarray}
By defining the RPA vacuum as 
\begin{eqnarray}
Q_{\mu i} \RPAr = 0,
\end{eqnarray}
and the RPA excited state as
\begin{eqnarray}
Q_{\mu i}^\dagger \RPAr  = |\omega_{\mu i}\rangle,
\end{eqnarray}
the  orthogonality condition of Eq. \ref{RPA-orth} is 
fulfilled automatically for single RPA excitations
\begin{eqnarray}
\RPAl Q_{\mu i} Q_{\mu' i'}^\dagger\RPAr 
&=&  \RPAl [ Q_{\mu i},Q_{\mu' i'}^\dagger]\RPAr \nonumber \\
&=&  
\delta_{\mu\mu'} \delta_{ii'}.
\end{eqnarray}

The RPA excitations are true oscillations of electronic systems in the
sense that the Heisenberg equations of motion for the excitation creation
 operators
is precisely as we expect for a harmonic system,
\begin{eqnarray}
[H_2,Q_{\mu i}^\dagger] = Q_{\mu i}^\dagger \omega_i,
\label{hberg}
\end{eqnarray}
where $\omega_i$ is the RPA frequency.  
From this we can
derive the RPA amplitudes, $\X^i_{ph}$ and $\Y^i_{ph}$,
and RPA frequencies $\omega_{i}$. After a bit
of algebra we arrive at the equations:
\begin{eqnarray}
E_{ph}\X^{i}_{ph} + \sum_{p'h'}(pp'|v|h'h)\Y^{i}_{p'h'} &-&
\sum_{h'p'}(ph'|v|p'h)\X^{i}_{p'h'}\nonumber \\
 &=& \omega_i \X^i_{ph} ,\\
E_{ph}\Y^{i}(ph) + \sum_{p'h'}(pp'|v|h'h)\X^{i}_{p'h'} &-&
\sum_{h'p'}(ph'|v|p'h)\Y^{i}_{p'h'}\nonumber \\
 &=&- \omega_i \Y^i_{ph}.
\end{eqnarray}
In matrix form:
\begin{eqnarray} 
\left(
\begin{array}{cc}
{\sf A} & {\sf B} \\
{\sf B} & {\sf A} 
\end{array}
\right)
\left(
\begin{array}{c}
|\X\> \\ |\Y \>
\end{array}
\right) = \omega
\left(
\begin{array}{c}
|\X \>
\\ 
-|\Y \>
\end{array} 
\right).
%$$\nonumber \\
%$${\sf M}\left(
%$$\begin{array}{c}
%$$|\X\>
%$$ \\ 
%$$|\Y \>
%$$\end{array} 
%$$\right) = \omega\left(
%$$\begin{array}{c}
%$$|\X \>\\ 
%$$-|\Y\> 
%$$\end{array} 
%$$\right).
\end{eqnarray}
The diagonal blocks,  ${\sf A}$, 
 are determined by Coulombic interactions between
single particle/hole excitations. 
\begin{eqnarray}
A_{php'h'} = E_{ph}\delta_{pp'}\delta_{hh'} - (ph'|v|p'h).
\end{eqnarray}
However, the off diagonal terms,
\begin{eqnarray}
B_{php'h'} =  (pp'|v|h'h),
\end{eqnarray}
are related fluctuations about the HF vacuum and thus give rise
to a net polarization of the HF vacuum.  Thus, the RPA vacuum differs
from the HF vacuum by the addition of these polarization
fluctuations. Lastly, we note, that if we take $\Y \rightarrow 0$ or
assume that ${\sf B}\approx 0$ and neglect the fluctuation terms, the
RPA equations reduce to the Tamm-Dankoff or configuration interaction
equations with single excitations.

%%%%%%%%%%%%%%%%%%%%%%%%%%%%%%%%%%%%%%%%%%%%%%%%%%%%%%%%%%%%%%%%%%%
%%%               connection with TD-DFT
%%%%%%%%%%%%%%%%%%%%%%%%%%%%%%%%%%%%%%%%%%%%%%%%%%%%%%%%%%%%%%%%%%
The matrix RPA method can be also derived as the small amplitude limit of
the time dependent HF equation \cite{Kerman76,Rowe}. We emphasize that the
time-dependent density functional theory \cite{Gross,Ahlrichs} is
equivalent to matrix RPA if we also assume harmonic oscillation of the
time-dependent density $\rho(t)$ around equilibrium. The standard
practical realization of the time-dependent density functional theory
is the time-dependent local density approximation \cite{Yabana98} where
nonlocal exchange interaction is approximated by local density dependent
interaction in adiabatic limit. In this respect the matrix RPA method we
use is advantageous because this method can treat nonlocal Fock
interactions in natural way.
%%%%%%%%%%%%%%%%%%%%%%%%%%%%%%%%%%%%%%%%%%%%%%%%%%%%%%%%%%%%%%%%%%%%

We can transform this into an eigenvalue problem by multiplying the
right hand side of this equation by the Pauli spin matrix, $\sigma_z$,
\begin{eqnarray}
{\sf M}| \Psi\>= \omega\sigma_z|\Psi\>,
\end{eqnarray}
where 
\begin{eqnarray}
|\Psi\>= \left(
\begin{array}{c}
|\X\> \\ 
|\Y \>
\end{array} 
\right),
\end{eqnarray}
is a two component spinor state.
The  eigenvector solution to this equation can be obtained by
transforming the RPA matrix equation
\begin{eqnarray}
\sigma_z^T{\sf M}|\Psi\> &=& \omega |\Psi \>.
\end{eqnarray}
Thus,  $|\Psi\>$ is the right-handed eigenvector of the antisymmetric matrix
\begin{eqnarray}
\sigma_z^T{\sf M}=
\left(
\begin{array}{cc}
{\sf A} & {\sf B} \\
-{\sf B} & -{\sf A} 
\end{array}
\right).
\end{eqnarray}
The RPA eigenstates are subject to the orthogonality condition
\begin{eqnarray}
\sum_{ph}{\X^i_{ph}}^* \X^j_{ph}  -
{\Y^i_{ph}}^{*} \Y^j_{ph}  = \delta_{ij}.
\end{eqnarray}
Thus, we define $\Psi^R$ as the right-handed eigenvector
\begin{eqnarray}
|\Psi^R_i\> = \left(
\begin{array}{c}
|\X^i\> \\ |\Y^i\>
\end{array}\right),
\end{eqnarray}
and $\Psi^L$ as the left-handed eigenvector
\begin{eqnarray}
|\Psi^L_i \>= \left(
\begin{array}{c}
|\X^i\> \\ -|\Y^i\>
\end{array}\right) = \sigma_z |\Psi_i^R\>.
\end{eqnarray}
The inner product relation between the left and right-handed
eigenvectors produces the desired orthogonality relation for the RPA
states. 
\begin{eqnarray}
\<\Psi^L_i|\Psi^R_j\> = \Psi^L_i \cdot \Psi^R_i  = \delta_{ij}.
\end{eqnarray}
These states are determined at the start of the calculation in a basis of
single particle/hole states which spans the  Fock space. 

The RPA frequency, $\omega_i$, is determined by taking the expectation
value of ${\sf M}$ between the left and right-handed eigenvectors. 
\begin{eqnarray}
\omega_i[\X,\Y,\phi,{\bf R}]  
&=& \<\Psi_i^L|{\sf M}|\Psi_i^{R}\> \nonumber \\
&=& 
\left(\X_k^*,-\Y_k^*\right)
\left(
\begin{array}{cc}
{\sf A} & {\sf B} \\
-{\sf B} & -{\sf A} 
\end{array}
\right)
\left(
\begin{array}{c}
\X_k \\ \Y_k 
\end{array}
\right)
%\nonumber \\
%&=&
%\sum_{ph}
%E_{ph}(|\X^k_{ph}|^2 +  |\Y^k_{ph}|^2 ) \nonumber \\
%&-&
%\sum_{php'h'}
% (ph'|v|p'h)(\X^{k*}_{ph}\X^k_{p'h'} - \Y^{k*}_{ph}\Y^{k*}_{p'h'}) \nonumber \\
%&+&\sum_{php'h'}(pp'|v|h'h) (\X^{k*}_{ph}\Y^{k}_{p'h'} +\Y^{k*}_{ph}\X^k_{p'h'}).
\end{eqnarray}

%%%%%%%%%%%%%%%%%%%%%%%%%%%%%%%%%%%%%%%%%%%%%%%%%%%%%%%%%%%%%%%%%%%%%%
\subsection{Equations of motion for the excited states molecular
dynamics in the RPA}
%%%%%%%%%%%%%%%%%%%%%%%%%%%%%%%%%%%%%%%%%%%%%%%%%%%%%%%%%%%%%%%%%%%%%%

The energy functional for the RPA equations is
\begin{eqnarray}
E_{ex}(i)&=&  \RPAl Q_{\mu i} H Q_{\mu i}^\dagger\RPAr ,\nonumber \\
&=& \RPAl H \RPAr \nonumber \\ &+&
\RPAl \left[Q_{\mu i},\left[ H, Q_{\mu i}^\dagger \right]
 \right]\RPAr,\nonumber  \\
&=&E_{o}+\omega_{i},
\end{eqnarray}
where $E_{o}$ is the RPA ground state energy and 
\begin{eqnarray}
\omega_i = 
\RPAl \left[Q_{\mu i},\left[ H, Q_{\mu i}^\dagger\right]\right]\RPAr,
\label{rpafreq}
\end{eqnarray}
is the RPA  frequency.  
In the previous section the RPA equation is derived via the 
Heisenberg equation  (Eq.\ref{hberg}) of motion for the 
excitation creation operators.
The Heisenberg equation  is physically 
the same as the requirement of a variational minimum
for the RPA excited state. To demonstrate this, we multiply
from the left with an arbitrary 
variation of the RPA excited state:
\begin{eqnarray}
 \RPAl \delta Q_{\mu i} \left[ H, Q_{\mu i}^\dagger\right] \RPAr
&=&\omega_{i} \RPAl \delta Q_{\mu i} Q^{\dagger}_{\mu i}\RPAr. \nonumber 
\\
\end{eqnarray}
We can rewrite this equation as a variational of the double commutator
in Eq.~\ref{rpafreq},
because $\RPAl  Q^{\dagger}_{\mu i} =0 $
\begin{eqnarray}
\frac{\delta}{\delta Q_{\mu i}}
\left \{
\RPAl \left[Q_{\mu i}, \left[ H, Q_{\mu i}^\dagger\right]
\right] \RPAr 
-\omega_{i}( \RPAl Q_{\mu i} Q^{\dagger}_{\mu i}\RPAr -1)
\right\}
\label{EoM}
=0
\end{eqnarray}
where $\omega_{i} $ plays a role of the Lagrange multiplier 
and insures the normalization of the RPA excited states.

%%%%%%%%%%%%%%%%%%%%%%%%%%%%%%%%%%%%%%%%%%%%%%%%%%%%%%%%%%%%%%%%%%%
%%%%%%%%%  connection with existing RPA literature
%%%%%%%%%%%%%%%%%%%%%%%%%%%%%%%%%%%%%%%%%%%%%%%%%%%%%%%%%%%%%%%%%%%
Neglecting vacuum polarization effects, i.e.
approximating $\RPAr$ by the HF Slater determinant
$|0_{HF}\>$  in Eq. \ref{EoM} we obtain
 the so-called equation of motion method 
which was originally proposed by D.J. Rowe \cite{Rowe}
and widely used in the studies of molecular excited states
in quantum chemistry \cite{Yeager75,Lasaga79,Ito96}.
The same matrix RPA equation can be derived by several methods,
e.g. by Green function methods \cite{Negele,Baldo77}, linear response
function method \cite{Olsen85,Ring} 
Each of these methods has
the strong points in specific aspects but is deficient in 
other respects.
As we demonstrated by Eq.\ref{EoM} our method is equivalent to
the Rayleigh - Ritz  variational principle for the 
RPA excitation amplitudes and the use of the variational approach
is pivotal to formulation classical equations of motion for the dynamical
optimization of the excited states. 
%%%%%%%%%%%%%%%%%%%%%%%%%%%%%%%%%%%%%%%%%%%%%%%%%%%%%%%%%%%%%%%%%%%%%%%%%

Finally, we assume that polarization of the ground state vacuum due to
particle-hole interaction is weak, then the RPA vacuum is
approximately the same as the HF or KS vacuum and we can write the RPA
ground state energy as $E_{o}\approx E_{HF}$.

Following these considerations, we can write the Lagrangian 
\begin{eqnarray}
\L &=& \half\sum_i\nu\int d1 |\dot{\phi}(1)|^2
+ \half\sum_{k}\mu \sum_{ph}|\dot{\X}^k_{ph}|^2 \nonumber \\
&+& \half\sum_{k}\mu \sum_{ph}|\dot{\Y}^k_{ph}|^2
+ \half\sum_n m_n \dot{{\bf R}}_n^2 \nonumber \\
&-&E_{HF}[\phi,{\bf R}]-\omega_{i}[ \phi,\X,\Y,{\bf R}]  \nonumber \\
&+& \sum_{ij}\Lambda_{ij}(\int d1 \phi^*_i(1)\phi_j(1) -\delta_{ij}) 
\nonumber \\
&+& \sum_{kl}\Gamma_{kl}(\<\Psi_k|\Psi_l\> -\delta_{ij}) 
\end{eqnarray}
where the constraint
\begin{eqnarray}
\sum_{kl}\Gamma_{kl}(\<\Psi_k&|&\Psi_l\> -\delta_{kl}) 
\nonumber \\
&=& \sum_{kl}\Gamma_{kl}\left(\sum_{ph}\X^k_{ph}{\X^l_{ph}}^{*}
-\Y^k_{ph}{\Y^l_{ph}}^{*} -\delta_{kl}\right)\nonumber \\
\end{eqnarray}
insures the orthogonality of the RPA excited states.
Solving the Euler-Lagrange equations for the single particle
amplitudes, $\phi_k(1)$,
\begin{eqnarray}
\nu\ddot{\phi}_i(1) &=& -\frac{\delta E_{HF}[\phi,{\bf R}]}{\delta
\phi_i^*(1)} 
-\frac{\delta \omega_{i}[ \phi,\X,\Y,{\bf R}]}{\delta \phi_i^*(1)} 
\nonumber \\
&+& \sum_j \Lambda_{ij}\phi_j(1)\label{eq:rpa1}
\end{eqnarray}
The equations for the single particle orbitals are similar to the 
CP equations written above (Eq.~\ref{eq:CP}) and include a term due to
the electronic excitations.  
In the case of a collective excitation, in which the RPA excitation is
delocalized over a number of single particle states
(or in other worlds, the oscillator strength of the RPA excitations is
 distributed more or less uniformly over a number of particle-hole excitations) 
the variation of the energy of collective excitation
$\omega$ with an infinitesimal change in one
of the single particle orbitals will be very small. Because of this,
the RPA excitation variables will evolve on a slower time-scale than the single
particle variables and we can
invoke an adiabatic separation between the RPA variables and the
single particle states. That is to say
\begin{eqnarray}
\frac{\delta \omega_k}{\delta \phi_j} \ll \frac{\delta
E_{HF}}{\delta \phi_j},
\label{eq:collective}
\end{eqnarray}
so that if one deals with collective electron 
motions, Eq.~\ref{eq:rpa1} can be approximated 
with very high degree of accuracy by the CP equations.
\begin{eqnarray}
\nu\ddot{\phi}_i(1) &=& -\frac{\delta E_{HF}[\phi,{\bf R}]}{\delta
\phi_i^*(1)} 
+ \sum_j \Lambda_{ij}\phi_j(1).
\end{eqnarray} 
This should be a reasonable approximation for many real molecular systems
where one has delocalized valence electrons  and relatively strong residual
Coulomb interaction between them.
We emphasize that our computational scheme is not restricted 
by the case of collective electronic excitation, i.e.
by the approximation (Eq.~\ref{eq:collective}).
Taking into account the explicit expression \cite{Ortiz94} for 
the functional derivatives of $\omega_k$ in respect to the 
molecular orbitals $\phi_i$
our dynamical equations can be straightforwardly used to
do molecular dynamics on ``non-collective'' excited state surfaces.

The dynamical equations for the RPA amplitudes are
determined to be
(assuming only a single RPA excitation in the system),
\begin{eqnarray}
\mu\ddot{\X}^k_{ph}& = &
-\frac{\delta \omega_{k}[\phi,\X,\Y,{\bf R}]}{\delta \X^{k*}_{ph}}
+ \Gamma_{kk} \X^k_{ph} \nonumber \\
&=&E_{ph}\X^k_{ph}
+\sum_{p'h'}((pp'|v|h'h)\Y^k_{p'h'}-(ph'|v|p'h)\X^k_{p'h'})\nonumber \\
&+& \Gamma_{kk} \X^k_{ph},
\label{rpa-mot1}
\end{eqnarray}
\begin{eqnarray}
\mu\ddot{\Y}^k_{ph} &=& 
-\frac{\delta \omega_{k}[\phi,\X,\Y,{\bf R}]}{\delta \Y^{k*}_{ph}}
- \Gamma_{kk} \Y^k_{ph} \nonumber \\
&=& 
E_{ph} \Y_{ph}^k + \sum_{p'h'} \left( (p p'|v|h' h)\X^k_{p'h'}
-(p h' |v|p' h)\Y^k_{p'h'} \right)
\nonumber \\ &-& \Gamma_{kk}\Y^k_{ph}.
\label{rpa-mot2}
\end{eqnarray}
or in matrix form

\begin{eqnarray} 
\mu
\left(
\begin{array}{c}
\ddot{|\X\>} \\ \ddot{|\Y \>}
\end{array}
\right)
=
\left(
\begin{array}{cc}
{\sf A}+ {\sf \Gamma}  & {\sf B} \\
{\sf B} & {\sf A} - {\sf \Gamma} 
\end{array}
\right)
\left(
\begin{array}{c}
|\X\> \\ |\Y \>
\end{array}
\right)
\end{eqnarray}

One important point to consider is that the functional derivatives of
$E_{HF}$ with respect to the particle states (i.e. single particle states
above the Fermi energy)vanish since the HF density matrix does not
include these states.  As we seen below, we need these states and
their energies to compute the Coulomb matrix elements in the RPA
dynamical equations.   However, all is not lost since we have at hand
the states below the Fermi energy and hence the ground state density.
We can thus reconstruct and simply diagonalize $H_1[\rho]$ to obtain
the single particle excited states.

Finally for the nuclear positions we have classical equations of
motion on the excited state potential energy surface, 
\begin{eqnarray}
\mu\ddot{{\bf R}}_n &=& 
-\frac{\delta E_{ex}[\phi,\X,\Y,{\bf R}]}{\delta {\bf R}_n}  \nonumber \\
&=& 
-\frac{\delta E_{HF}[\phi,{\bf R}_n]}{\delta {\bf R}_n} 
-\frac{\delta \omega_{k}[\phi,\X,\Y,{\bf R}_n]}{\delta {\bf R}_n} \nonumber 
\end{eqnarray}
The details of the derivation of each of the functional
derivatives above are provided in the Appendix A. The derivation of the
molecular dynamics 
equations of motion in the RPA
compose the central result of this paper.

\section{Numerical and Analytical Examples}

%%%%%%%%%%%%%%%%%%%%%%%%%%%%%%%%%%%%%%%%%%%%%%%%%%%%%%%%%%%%%%%%%%%%
\subsection{The dynamical optimization of
the excited state on a two level model}
%%%%%%%%%%%%%%%%%%%%%%%%%%%%%%%%%%%%%%%%%%%%%%%%%%%%%%%%%%%%%%%%%%%%

In order to get an indication of how our approach works, we apply our
scheme to a model two-level system consisting of $N$ {\em
distinguishable} electrons labeled by $p=1,2,...,N$. Each of these can
occupy one of two orbitals, a lower and upper, having energies
$\frac{1}{2}\varepsilon(R)$ and $-\frac{1}{2}\varepsilon(R)$ and
distinguished by $g=1$ and $g=2$ respectively.  The coupling constant
$V(R)$ does not depend on any quantum number.  The gap between orbital
$\varepsilon(R)$ and the strength of interaction $V(R)$ depends on a
classical variable $R$.  The HF approximation is obtained by setting
$V=0$, such that the lower level is fully occupied and the upper is
empty, i.e. $N = \Omega$ where $\Omega$ is the degeneracy of the
levels.  This model encapsulates the salient effects one expects to
see in physically realistic systems.  The model Hamiltonian possesses
the SU(2) symmetry \cite{Lipkin65} and can be written in terms of
quantum quasi-spin operators and the classical variable $R$:
\begin{eqnarray}
H&=& \varepsilon(R) \hat{J}_{z} -\frac{1}{2} V(R) 
\left(\hat{J}_{+} \hat{J}_{+} + \hat{J}_{-}\hat{J}_{-} \right) \nonumber \\
&+& \frac{M {\dot{R}}^2}{2} + \frac{1}{2} K (R-R_0)^2
\end{eqnarray}
where the operators 
 $ \hat{J}_{+}, \hat{J}_{-}, \hat{J}_{z}$ are defined as follows:
\begin{eqnarray}
\hat{J}_{z}&=&\frac{1}{2} \sum_{p=1}^{\Omega}
\left( a^{\dagger}_{2 p} a_{2 p} - a^{\dagger}_{1 p} a_{1 p}\right) , 
\nonumber \\
\hat{J}_{+}&=&\sum_{p=1}^{\Omega} a^{\dagger}_{2 p}a_{1 p} \nonumber \\
\hat{J}_{-} &=& \left(\hat{J}_{+}\right)^{\dagger}.
\end{eqnarray}
Above $a^{\dagger}_{g p}, a_{g p}$  are particle creation and annihilation
operators on the lower  ($g=1$) or the upper ($g=2$) level.
$K$ is the spring constant which does not allow to classical system 
to move far from the equilibrium.   

Place the classical variable at the position
\begin{eqnarray}
R=R_0 +q ,
\end{eqnarray}
which is the sum of the equilibrium position $R_0$
and the displacement $q$. Assuming the displacement from equilibrium
is small, 
we can expand the Hamiltonian in $q$ 
\begin{eqnarray}
\varepsilon (R)= \varepsilon(R_0) + 
\left.\frac{d \varepsilon}{d R}\right|_{R_0} q+ O(q^2)
\label{varepsilon}
\\
V(R)= V(R_0) + \left.\frac{d V}{d R}\right|_{R_0} q+ O(q^2).
\end{eqnarray}
Truncating this at first order gives a linear interaction between
classical and quantum systems.  Neglecting the terms of $O(q^2)$ we
get the following quantum many-body Hamiltonian which depends
parametrically on the displacement $q$ of the classical variable:

\begin{eqnarray}
H(q)&=& \left(\varepsilon(R_0) + 
\left.\frac{d \varepsilon}{d R}\right|_{R_0} q\right) 
 \hat{J}_{z}  \nonumber \\
&-&\frac{1}{2} \left(V(R_0) + \left.\frac{d V}{d R}\right|_{R_0} q\right) 
\left(\hat{J}_{+} \hat{J}_{+}
+ \hat{J}_{-}\hat{J}_{-} \right) \nonumber \\
&+& \frac{M {\dot{q}}^2}{2} + \frac{1}{2} K q^2
\end{eqnarray}

In the RPA, the operators $\hat{J}_{+}, \hat{J}_{-}$ behave like 
pure bosonic operators
\begin{eqnarray}
\left[ \frac{\hat{J}_{-}}{\sqrt{N}}, \frac{\hat{J}_{+}}{\sqrt{N}}\right]
= 1
\end{eqnarray}
Using these bosons, we construct the RPA excitation creation operator
(in the two-level model there is only one possible RPA
excitation)

\begin{eqnarray}
Q^{\dagger}= \frac{1}{\sqrt{N}} 
( \X \hat{J}_{+} - \Y \hat{J}_{-} )
\end{eqnarray}

The orthogonality of the RPA excited  state imposes the 
following normalization 
condition on the RPA amplitudes:
\begin{eqnarray}
\X^2 -\Y^2=1
\end{eqnarray}
The RPA frequency is given by the matrix element
\begin{eqnarray}
\omega[\X,\Y,q]&=&
\RPAl \left[Q ,\left[ H(q),Q^{\dagger} \right] \right]\RPAr
\nonumber
\\
&=&
\left(\varepsilon(R_0)+ \left.\frac{d \varepsilon}{d R}\right|_{R_0} q\right)
(\X^{2}+\Y^{2}) \nonumber \\
&-&2 \left(V(R_0) + \left.\frac{d V}{d R}\right|_{R_0} q\right) N \X \Y
\end{eqnarray}
We can rewrite the $\omega[\X,\Y,q]$ as a functional of only the $\Y$ 
amplitudes 
and classical  $q$ variable:
\begin{eqnarray}
\omega[\Y,q]&=&
\left(\varepsilon(R_0)+ \left.\frac{d \varepsilon}{d R}\right|_{R_0}
q\right)
(1+2 \Y^{2}) \nonumber \\ &-&
2 \left(V(R_0) + \left.\frac{d V}{d R}\right|_{R_0} q\right)
 N\sqrt{1+\Y^2} \Y
\end{eqnarray}
The variational minimum of the RPA frequency can be determined
analytically and is given as a the function of $q$:
\begin{eqnarray}
\omega(q) &=&\sqrt{\left( \varepsilon(R_0)+ 
\left.\frac{d \varepsilon}{d R}\right|_{R_0} q \right)^2 
- \left( V(R_0) + \left.\frac{d V}{d R}\right|_{R_0} q \right)^2
N^2}.\nonumber \\
\label{omega}
\end{eqnarray}

The classical Lagrangian for the annealing of the 
electronic excited state is given by
\begin{eqnarray}
L[q,Y]= \frac{1}{2}\mu \dot{\Y}^{2}- E_{HF}[q]- 
 \omega[\Y,q]+ \frac{1}{2} m \dot{q}^{2}-  \frac{1}{2} K q^2 \; ,
\end{eqnarray}
where the $E_{HF}[q]$ is the HF ground state energy.

The Lagrangian produces the  equations of motion for the RPA amplitudes
\begin{eqnarray}
\mu \ddot{\Y}&=&
-4 \left(\varepsilon(R_0)+ \left.\frac{d \varepsilon}{d
R}\right|_{R_0} 
q\right) \Y  \nonumber \\ &+&
2 \left(V(R_0) + \left.\frac{d V}{d R}\right|_{R_0} q\right) 
N \frac{1+2 {\Y}^{2}}{\sqrt{1+ {\Y}^{2}}}
\\
m \ddot{q}&=&
\left.\frac{d \varepsilon}{d R}\right|_{R_0} (1+2 \Y^{2} +\frac{N}{2}) 
\nonumber \\ 
&-&2 \left.\frac{d V}{d R}\right|_{R_0}  N \sqrt{1+ {\Y}^{2}} \Y -  K q
\end{eqnarray}
The solution of these differential equations starting from an arbitrary value
of $Y(\tau=0)$ and $q(\tau=0)$ is shown in Fig.1.  The solutions oscillate
about the minimum so that in order to
get converged solution we introduced a
velocity damping term to suppress oscillation around the 
equilibrium configuration. 
The final value of the amplitudes $Y$ and $X=\sqrt{1+Y^2}$
are in excellent agreement with the analytical RPA results
for a given equilibrium value of the classical displacement 
$q_0= q(\tau \to \infty)$:
\begin{eqnarray}
Y&=&\sqrt{\frac{\varepsilon (R_0 +q_0) -\omega(q_0)}{2
\omega(q_0)}},\nonumber \\  
X&=&\sqrt{\frac{\varepsilon (R_0 +q_0) +\omega(q_0)}{2 \omega(q_0)}}
\end{eqnarray}
Here the function $\varepsilon (R_0 +q_0)$ is given by Eq.~\ref{varepsilon} and 
$\omega(q_0)$ by the Eq.~\ref{omega}.

The simplified form of the usual CPMD equation of motion can be
derived within our model if we set the residual interaction $V(R)=0$
and consider HF gap $\varepsilon(R)$ as a fixed parameter. In this
case only the classical degree of freedom evolve in time. One can see
from the lower part of the Fig.1 that there is a profound difference
between excited and ground state dynamics for a classical system.
This difference is  due to the Hellmann-Feynman force from the
RPA excited state (App. A.1).

%%%%%%%%%%%%%%%%%%%%%%%%%%%%%%%%%%%%%%%%%%%%%%%%%%%%%%%%%%%%%%%%%%%%%%%%%%

\subsection{Excited state dynamics in conjugated polymer lattices}

%%%%%%%%%%%%%%%%%%%%%%%%%%%%%%%%%%%%%%%%%%%%%%%%%%%%%%%%%%%%%%%%%%%%%%%%%%

In this example we take a more challenging case in which the ground 
state can be determined exactly, but the excited states cannot.  The 
${\pi}$-electrons in linear polyene systems, such as {\it 
trans}-poly-acetylene, can to a fair approximation be treated as a 
quasi-one dimensional electron gas within the tight-binding 
approximation.  Accordingly, we start our consideration from the 
following model electronic Hamiltonian for the polymer chain

\begin{eqnarray}
H  &=& H_o + : H_{int} : \\
H_o&=&\sum_{n,\sigma}({t_o}+\alpha(u_{n+1}-{u_n}))
(c_{n+1,\sigma}^{\dagger} c_{n,\sigma}
+c_{n,\sigma}^{\dagger} c_{n+1,\sigma}) \nonumber \\
&+&\frac{K}{2}\sum_n(u_{n+1}-u_n)^2 \\
H_{int} &=& \frac{1}{2} \sum_{nm} V(n,m) \sum_{\sigma \sigma'}
c^{\dagger}_{n \sigma} c^{\dagger}_{m \sigma'} 
c_{m \sigma'} c_{n \sigma }
\end{eqnarray}
Here $c_{m \sigma}$ creates an electron of spin $\sigma$ on site $m$.
The mean field Hamiltonian, $H_o$ is based upon the so called
Su-Schrieffer-Heeger (SSH)\cite{SSH} model for the band-structure of
{\em trans}-polyacetylene.  
The first term in the $H_o$ gives the
energy for an electron with spin $\sigma$ to hop between neighboring
$p$-orbitals. The strength of this term is modulated by linear
coupling to distortions in the polymer lattice away from evenly spaced
lattice positions. Finally, the last terms in $H_o$ gives the harmonic
interactions between lattice-sites arising from the $\sigma$-bonds
between neighboring lattice atoms. The $:H_{int}:$ is the normal
ordered, i.e. irreducible to one body operators, interaction between
valence electrons from the $p_z$-orbitals.

The Hamiltonian $H_o$ is quadratic form in 
fermion creation/annihilation operators. This quadratic form
can be exactly diagonalized by the canonical transformation:
\begin{eqnarray}
c_{n \sigma} &=& \frac{1}{\sqrt{N}} \sum_{k} 
\left(\beta^*_k +i(-)^n \alpha^*_k) a_{k \sigma +}
+ (\alpha^*_k -i(-)^n \beta^*_{k} ) a_{k \sigma -}
\right) \nonumber\\
c^{\dagger}_{n \sigma} &=& \frac{1}{\sqrt{N}} \sum_{k} 
\left(\beta_k -i(-)^n \alpha_k) a^{\dagger}_{k \sigma +}
+ (\alpha_k +i(-)^n \beta_{k} ) a^{\dagger}_{k \sigma -}
\nonumber
\right) 
\end{eqnarray}
This canonical transformations is equivalent 
to a Hartree-Fock diagonalization of the electronic Hamiltonian
and produces the uncorrelated mean field Hamiltonian:
\begin{eqnarray}
H_o=\sum_k E_k(a^{\dagger}_{k\sigma+} a_{k\sigma+}
-a^{\dagger}_{k\sigma-} a_{k\sigma-}  )+2NKu^2
\end{eqnarray}
where $E_k$ is the HF  energy of a quasi-particle, $a_{k\sigma\pm}$, with 
wavevector $k$ in the first reduced Brillouin zone of the system, 
${-\pi /2a < k \leq \pi /2a }$.
We also define + to correspond to 
states above the Fermi-energy (particles) and $-$ to those below
(holes). The single-particle energy,
$E_k=\sqrt{\epsilon_{k}^{2}+\Delta_{k}^{2}}$, is a function of 
the unperturbed band energy, $\epsilon_k=2 {t_o} \cos(k a) $, and the 
phonon-induced energy gap, $\Delta_k(u)=2 \alpha u \sin(k a)$. 
A plot of the Hartree-Fock ground state energy surface in shown in Fig
2c. 

In this paper we are primarily interested in the details of
computational algorithm and method. In order to make our consideration
as model independent as possible we take the following general form of
the residual interaction:
\begin{eqnarray}
H_{int} =\sum_{k k' q} \sum_{\sigma \sigma'}
V(q) a^{\dagger}_{k \sigma \pm} a^{\dagger}_{k' \sigma \pm}
a_{k-q \sigma \pm} a_{k'+q \sigma \pm} \; ,
\end{eqnarray}
where the interaction $V(q)$ is just a Fourier transformation of 
the physically meaningful interaction in coordinate representation 
(with $U = 0.01$ eV and 
$r_o=a$):
\begin{eqnarray}
V_{n,n+1} & = & \frac{U}{1+(a+2u)/r_{o}} \mbox{\,\,\,\rm for even $n$},
\nonumber \\
V_{n,n-1} & =& \frac{U}{1+(a-2u)/r_{o}} \mbox{\,\,\,\rm for odd $n$},
\nonumber \\
V_{nn} & = & U \mbox{\,\,\, for $n$ even or odd}
\label{eq:interaction}
\end{eqnarray}

Using the 2-body interaction of this form (with $U = 0.01$ eV and 
$r_o=a$) and the description of the ground state given above, we 
integrated the RPA equations (Eq.~\ref{rpa-mot1} and 
Eq.~\ref{rpa-mot2}) in $k$-space over the first reduced Brillouin 
zone using the Verlet algorithm~\cite{Verlet} for both the RPA 
amplitudes and classical coordinate.  In all of our runs, we monitored 
the orthogonality of the amplitudes and re-orthogonalized whenever the 
orthogonality strayed from unity by 1 part in $10^{6}$.  In the first 
example, we chose our initial excited state such that $X_{ph}$ 
corresponded to the first CI(singles) triplet state and $Y_{ph}=0$.  
In all of these calculations, the fictitious electronic mass was 
chosen to be $\mu$=400 a.u. and the classical mass to be that of a 
-CH- monomer ($m=26000 $a.u.).

In the first example, shown in Fig.2a, 
we started from purely random distribution of 
initial amplitudes for $X$ and $Y$ again with $u(0)=0.1$\AA. As 
before, we dampened the RPA velocities to attempt to cool the system 
down to the excited state Born-Oppenheimer energy. Notice, that the 
system rapidly finds the lowest energy configuration of the excited 
state surface and begins to oscillate on the BO surface after only a 
few periods. This demonstrates the utility of our method in 
determining very rapidly minimal energy configurations on the excited 
state energy surface.

In the next example, our initial excited state was taken to be an
eigenstate of the $A$ matrix (equivalent to the CI singles matrix)
with orbital angular momentum 1 (triplet state). Starting from a
lattice distortion of $u$=0.1\AA, we solved the full RPA equations of
motion with the inclusion of a term to dampen the RPA amplitude
velocities.  This helped to keep the electronic velocities in
sufficient check and prevented significant excursions from the excited
state Born-Oppenheimer energy surface. The results of this calculation
are shown in Fig. 2b. Here we have plotted the excitation
energy, $\omega$ as a function of $u$ as the RPA amplitudes are
quenched to their minimal values. Notice, however, that even though we
kept the electronic velocities fairly "cool", there is some
"fuzziness" to our energy surface. This is partially due to the fact
that our initial state was not an exact solution to the RPA equations
and to the fact that there is some dynamical coupling between the
motions of the amplitudes and the classical degrees of freedom.  Thus,
the RPA amplitudes acquire a small amount of velocity as they are
quenched to the final solutions.

A final comment regarding the computational efficiency of our approach
is in due order. First of all, in the calculations presented herein,
we made use of an analytic representation of the ground state orbitals
and energies. In general, one will need to determine these for each
new molecular configuration either by using integrating the ground
state CP equations or by converging the KS or HF equations. Assuming
we were handed a set of single particle orbitals, we can compare the
computational cost associated with each molecular dynamics step in the
RPA with the cost required to setup and diagonalize the CI singles
(CIS) matrix. In Fig.~\ref{fig:cputimes} we show the ratio of CPU
times required to perform a CIS calculation versus the effort required
to perform a similar sized single MD time-step using the RPA/MD method
developed here as run on an SGI Origin2000 (on a single 195 MHz R10000
processor). In each case, the RPA/MD results were between 2 to 10
times as fast per time-step as diagonalizing the CI matrix with
performance degrading with the size of the number of particle/hole
states used.  The CIS results reported are the CPU times required
for the construction and diagonalization of the CIS matrix (converging
only the lowest eigenstate) and does not include the additional effort
need to compute analytical gradients which are needed in order to
perform any sort of molecular dynamics on the excited state surface.
Finally, we note that the each RPA/MD step requires the construct of
both the {\sf A} and {\sf B} matrices whereas  CIS requires only the 
{\sf A} matrix. 

We also point out that the majority of the CPU effort per time-step 
went into the construction of the two-body interaction matrix 
elements.  These matrix elements, along with the single particle 
orbitals, must be in hand at every time step and are a necessary 
component of any many-body treatment of the excited states.  This 
effort scales as $\approx N_{k}^{5}$ due to the transformation between 
the quasi-particle basis and the lattice representation.  For larger 
calculations involving the combination of many particle/hole state, 
this part of the calculation dominates the effort.  Furthermore, in a 
full implementation of this scheme, one is required to diagonalize the 
HF or KS matrix to obtain the orbitals above the Fermi-energy.  This 
too, adds to the net effort as it scales as $\approx N_{k}^{3}$.  In a 
general plane-wave implementation of this methods, we can utilize 
efficient and scalable multi-dimensional fast Fourier transformation 
algorithms and matrix-diagonalization methods to facilitate these two 
phases of the calculation.

%%%%%%%%%%%%%%%%%%%%%%%%%%%%%%%%%%%%%%%%%%%%%%%%%%%%%%%%%%%%%%%%%%%%%%%%%
\section{Discussion}
%%%%%%%%%%%%%%%%%%%%%%%%%%%%%%%%%%%%%%%%%%%%%%%%%%%%%%%%%%%%%%%%%%%%%%%%%

In this paper, we have presented a theoretical scheme for doing
molecular dynamics simulations of electronically excited systems
within the HF(KS)+random phase approximation.  In the spirit of Car
and Parrinello, we treat the RPA amplitudes, the single particle
orbitals, and the nuclear coordinates as dynamical variables subject
to a set of holonomic constraints which enforce the orthogonality
relations of the RPA amplitudes and orbitals.

As mentioned in the introduction, the two primary computational cost
incurred in applying the CPMD scheme are enforcing the orthogonality
constraints between the single particle states and the small time-step
required to integrate the CP equations.  As mentioned above, the
constraint part of each iteration scales as $\sim M\times N^2$.  In
the present scheme, we have and additional set of constraints for the
orthogonality of the RPA states. Likewise, the evaluation of the RPA
constraint should scale as $\sim M_{rpa}\times N_{RPA}^2$ where
$M_{rpa}$ is the number of particle-hole basis states used to
construct the RPA excitation and $N_{RPA}$ the number of RPA
excitations in the system.  Since we are primarily interested in
systems with one RPA excitations and we can restrict the excitation to
a limited number of particle-hole states, the added constraint does
not pose a substantial increase in the computational
effort. Furthermore, the orthogonality between the RPA state and the
ground state is implicit in the construction of the RPA state and,
hence, imposes no additional difficulties. Note, that this would not
be the case had we chosen a perturbative construction of the excited
state.

A sample calculation on an albeit simple model system produces exact
agreement with analytical results.  While this model avoids a number
of computational difficulties in using this approach, it does give an
indication of how the equations of motion can be applied. 
%
%%%%%%%%%%%%%%%%%%%%%%%%%%%%%%%%%%%%%%%%%%%%%%%%%%%%%%%%%%%%%%%%%%%
The more realistic simulations of the excited state dynamics 
of one-dimension polymer chains demonstrate that our proposed 
approach is much more computationally advantageous since it
considerably decreases the computational efforts over the direct 
solution of the RPA eigen-problem.
%%%%%%%%%%%%%%%%%%%%%%%%%%%%%%%%%%%%%%%%%%%%%%%%%%%%%%%%%%%%%%%%%%%%
%
We are
currently developing a much more realistic and generalizable
application of our excited states molecular dynamics for simulating
the photo-excitation dynamics in $\pi$-conjugated systems such as
conducting polymers, ~\cite{ERBDSK2} as well as the inclusion of
transitions between RPA states, such as non-radiative triplet-singlet
transitions.

%%%%%%%%%%%%%%%%%%%%%%%%%%%%%%%%%%%%%%%%%%%%%%%%%%%%%%%%%%%%%%%%%%%%
\acknowledgments
%%%%%%%%%%%%%%%%%%%%%%%%%%%%%%%%%%%%%%%%%%%%%%%%%%%%%%%%%%%%%%%%%%%%
 
This work was supported in part by the R. A. Welch Foundation (E-1337)
and by the National Science Foundation (CHE-9713681).

%%%%%%%%%%%%%%%%%%%%%%%%%%%%%%%%%%%%%%%%%%%%%%%%%%%%%%%%%%%%%%%%%%%%
\appendix
%%%%%%%%%%%%%%%%%%%%%%%%%%%%%%%%%%%%%%%%%%%%%%%%%%%%%%%%%%%%%%%%%%%%

%%%%%%%%%%%%%%%%%%%%%%%%%%%%%%%%%%%%%%%%%%%%%%%%%%%%%%%%%%%%%%%%%%%%
\section{Evaluation of Functional Derivatives}
%%%%%%%%%%%%%%%%%%%%%%%%%%%%%%%%%%%%%%%%%%%%%%%%%%%%%%%%%%%%%%%%%%%%

In this Appendix we compute the various functional derivatives of the
RPA energy functional needed to solve the dynamical RPA equations
presented above.  
\subsection{Hellmann-Feynman forces for excited states}
We first invoke the Hellmann-Feynman
theorem~\cite{Hellmann37,Feynman39}
to provide the forces for
the evolution of the nuclei on the Born-Oppenheimer energy surface of
the RPA excitations.
\begin{eqnarray}
\frac{\delta E_{ex}[\phi,\X,\Y,{\bf R}]}{\delta {\bf R}}=
\frac{\delta \RPAl Q_{\mu i} H Q_{\mu i}^\dagger\RPAr }
{\delta {\bf R}}
\end{eqnarray}
The Hellmann-Feynman theorem is indeed valid for
the RPA excited states, since
\begin{eqnarray}
\frac{\delta \RPAl Q_{\mu i} Q_{\mu i}^\dagger\RPAr}{\delta {\bf R}}=
\frac{\delta \RPAl \left[Q_{\mu i}, Q_{\mu i}^\dagger\right]\RPAr}
{\delta {\bf R}}=0
\end{eqnarray}
Thus,
\begin{eqnarray}
\frac{\delta E_{ex}[\phi,\X,\Y,{\bf R}]}{\delta {\bf R}}=
 \RPAl Q_{\mu i}\frac{\partial H}{\partial {\bf R} }Q_{\mu i}^\dagger\RPAr
\end{eqnarray}
To calculate this matrix element we first rewrite the gradient of 
$H_2$ with respect to the nuclear coordinates 
in a second quantized form in the basis of
electron creation and annihilation operators:
\begin{eqnarray}
\frac{\partial H}{\partial {\bf R}}= 
\sum_{ i j} (i|\frac{\partial H}{\partial {\bf R}}|j)
\sum_{\sigma} a^{\dagger}_{i \sigma} a_{j \sigma}.
\end{eqnarray}
Here the matrix element 
\begin{eqnarray}
(i|\frac{\partial H}{\partial {\bf R}}|j)=
\int dr
 \phi^*_i (r) \frac{\partial H}{\partial {\bf R}} (R,r) \phi_j (r)
\end{eqnarray}

Using this, one can easily show
\begin{eqnarray}
\RPAl Q_{\mu i}\frac{\partial H}{\partial {\bf R} }Q_{\mu i}^\dagger\RPAr &=& 
 \RPAl  \frac{\partial H}{\partial {\bf R}}\RPAr \nonumber \\ 
&+&
\RPAl \left[ Q_{\mu i}, 
\left[ \frac{\partial H}{\partial{\bf R}}, Q_{\mu i}^\dagger
\right] \right] \RPAr
\end{eqnarray}
Neglecting the polarization fluctuation in HF vacuum,
we arrive at the following expression for the classical forces:
\begin{eqnarray}
\RPAl Q_{\mu i}\frac{\partial H}{\partial {\bf R}}Q_{\mu i}^\dagger\RPAr
&=&
2 \sum_{hh}(h| \frac{\partial H}{\partial {\bf R}} |h)\nonumber \\
&+&\sum_{php'} (p'| \frac{\partial H}{\partial{\bf R}} |p)
\left( X^{i*}_{ph} X^i_{p'h} +Y^{i*}_{ph} Y^{i}_{p'h} \right)
\nonumber
\\
&+&
\sum_{phh'} (h| \frac{\partial H}{\partial {\bf R}} |h')
\left( X^{i*}_{ph} X^i_{ph'} +Y^{i*}_{ph} Y^{i}_{ph'}\right)\nonumber \\
\end{eqnarray}
where the first term is simply the ground state Hellmann-Feynman force
and the remaining terms are due to the electronic excitations.
\subsection{Functional derivatives for the amplitude motions.}
The functional derivatives for the evolution of the RPA state and HF
vacuum are straightforward to compute and are presented below.
\subsubsection{
The functional derivative of the HF energy with respect to
the single particle/hole states.}
\begin{eqnarray}
\frac{\delta E_{HF}[\phi,{\bf R}]}{\delta\phi_i(1)}  &= &
\frac{\delta}{\delta\phi_i(1)}  \sum_j^{occ}\int d2d3
\phi_j^*(2)\phi_j(3) H(23;\rho) \nonumber \\
&=&
 \int d2\phi_i^*(2)H(21;\rho) \nonumber \\
&+&\left(\int d4 \phi_i^*(4)\right)
\sum_j^{occ}
\< \phi_j| \frac{\delta H[\rho]}{\delta\rho} |\phi_j\>
\end{eqnarray}

%%%%%%%%%%%%%%%%%%%%%%%%%%%%%%%%%%%%%%%%%%%%%%%%%%%%%%%%%%%%%%%%%%%%
\subsubsection{Functional derivatives for the RPA amplitudes}
%%%%%%%%%%%%%%%%%%%%%%%%%%%%%%%%%%%%%%%%%%%%%%%%%%%%%%%%%%%%%%%%%%%%%
First, in vector form
\begin{eqnarray}
\frac{\delta \omega_k}{\delta \<\X_k|} = 
{\sf A}|\X_k\> + {\sf B}|\Y_k\> 
\end{eqnarray}
and 
\begin{eqnarray}
\frac{\delta \omega_k}{\delta \<\Y_k|} = 
{\sf B}|\X_k\> + {\sf A}|\Y_k\>.
\end{eqnarray}
Written in terms of the components:
\begin{eqnarray}
\frac{\delta \omega_{k}[\phi,\X,\Y,{\bf R}]}{\delta \X^{k*}_{ph}}  &= &
\sum_{p'h'}(A_{php'h'}\X^k_{p'h'} + B_{php'h'}\Y^{k}_{ph} )
\end{eqnarray}
\begin{eqnarray}
\frac{\delta \omega_{k}[\phi,\X,\Y,{\bf R}]}{\delta \Y^{k*}_{ph}}  = 
\sum_{p'h'}(A_{php'h'}\Y^k_{p'h'} + B_{php'h'}\X^{k}_{ph} )
\end{eqnarray}

%%%%%%%%%%%%%%%%%%%%%%%%%%%%%%%%%%%%%%%%%%%%%%%%%%%%%%%%%%%%%%%%%%%%%
\section{Dynamical calculation of the Lagrange multipliers 
for the RPA excited states}
%%%%%%%%%%%%%%%%%%%%%%%%%%%%%%%%%%%%%%%%%%%%%%%%%%%%%%%%%%%%%%%%%%%%%

The evolution of excited states is subject to holonomic constraints
which enforce the orthogonality of the RPA amplitudes,
( Eq.~\ref{RPA-orth}).  Because of this, we have constraint 
forces $-\Gamma_{kk} X^{k}_{ph}$ and  $\Gamma_{kk} Y^{k}_{ph}$
in the dynamical equations(Eq.~\ref{rpa-mot1} and Eq.~\ref{rpa-mot2})
which keep the RPA amplitudes on the constrained surface.
However, the Lagrange multiplier $\Gamma$ is unknown. 

Denote by
\begin{eqnarray}
\sigma_{k}=
\sum_{ph}{\X^{k*}_{ph}} \X^{k}_{ph}
-{\Y^{k*}_{ph}} \Y^{k}_{ph} -1.
\end{eqnarray}
Since $\sigma_{k}$ is  the integral of motion
for the differential equations (Eq.\ref{rpa-mot1} and Eq.~\ref{rpa-mot2}),
the first and second time derivatives of $ \sigma_{k}$
should be zero.
\begin{eqnarray}
\ddot{\sigma}_{k}&=& \left(
2 \sum_{ph} {\dot{\X}^{k*}_{ph}} \dot{\X}^{k}_{ph}
-\dot{\Y}^{k*}_{ph} \dot{\Y}^{k}_{ph}  \right)
\nonumber
\\
&+&\sum_{ph}\left( \ddot{\X}^{k*}_{ph} \X^{k}_{ph}
+\X^{k*}_{ph} \ddot{\X}^{k}_{ph}
-\ddot{\Y}^{k*}_{ph} \Y^{k}_{ph} 
-\Y^{k*}_{ph} {\ddot\Y}^{k}_{ph}
\right)
\end{eqnarray}
We  can then eliminate the ``amplitudes accelerations'' 
$\ddot{\Y}^{k},\ddot{\X}^{k}$, ${\ddot{\Y}^{k*}},{\ddot{\X}^{k*}}$ 
by the use of the RPA equation of motion.
From these considerations, we obtain the 
expression for the Lagrange multiplier $\Gamma_{kk}$:
\begin{eqnarray}
\Gamma_{kk}&=& - \frac{1}{2 \sum_{ph} 
\X^{k*}_{ph} \X^{k}_{ph}
+\Y^{k*}_{ph} \Y^{k}_{ph}
} 
\nonumber \\
&\times&\sum_{ph} \left[ 
2\mu ( \dot{\X}^{k*}_{ph} \dot{\X}^{k}_{ph}
- \dot{\Y}^{k*}_{ph} \dot{\Y}^{k}_{ph}  )
\right. \nonumber\\
&-&\left. \frac{\delta \omega}{\delta X^{k}_{ph}} \X^{k}_{ph}
-\frac{\delta \omega}{\delta X^{k*}_{ph}} \X^{k*}_{ph}
+ \frac{\delta \omega}{\delta Y^{k}_{ph}} \Y^{k}_{ph}
+\frac{\delta \omega}{\delta Y^{k*}_{ph}} \Y^{k*}_{ph}
\right]\nonumber \\
\end{eqnarray}

%%%%%%%%%%%%%%%%%%%%%%%%%%%%%%%%%%%%%%%%%%%%%%%%%%%%%%%%%%%%%%%%%%%%%%
%\bibliography{Bibliography}
%%%%%%%%%%%%%%%%%%%%%%%%%%%%%%%%%%%%%%%%%%%%%%%%%%%%%%%%%%%%%%%%%%%%%%

%\bibliographystyle{jcp}

\newpage
\begin{center}
{\bf Figure Captions}
\end{center}

\begin{figure}
\caption[]{The evolution of the RPA amplitude $Y$ (a.)  and the
displacement $q$ (b.) in the classically-extended-SU(2) model.  The
dashed line is for simplified ground state CPMD; the result of our
approach is shown by solid line.  The following set of parameters is
used in this calculation: $ \mu=M=1$,$ \epsilon=1$, $ V=0.06 $,
$N=10$, $(d\epsilon/dr)_{R_0}=0.1$, $ (dV/dR)_{R_0}=0.06 $, $ K=10$.}
\label{fig:Y}
\end{figure}

\begin{figure}

\caption[]{{\bf Energy surfaces for ground and excited states for
linear polyene model.}  (a.) Excited state molecular dynamics starting
from a random RPA vector converging to the excited state
Born-Oppenheimer energy surface.  The excited state energy surface is
the sum of the excitation energy and the ground state surface plotted
in c.  Here, only the energy difference is plotted.  (b.) Excited
state molecular dynamics starting from the lowest triplet CI(S)
eigenstate. As in a., only the energy difference between the triplet
and ground state is shown.  (c.) Ground state Born-Oppenheimer
surface.  Note the change in energy scale between the ground and
excited state plots. See text for details. }
\label{fig2}
\end{figure}

\begin{figure}
\caption[]{Comparison of CPU effort for RPA/MD 
%%%%%%%%%%%%%%%%%%%%%%%%%%%%%%%%%%%%%%%%%%%%%%%%
%!!!!!!!!!!!!!!!!!!!!!!!!!!!!!!!!!!!!!!!!!!!!!!
%(with the ``backgoing'' amplitude $Y=0$)
%%%%%%%%%%%%%%%%%%%%%%%%%%%%%%%%%%%%%%%%%%%%%%%%
%%%%%%%%%%%%%%%%%%%%%%%%%%%%%%%%%%%%%%%%%%%%%%%
vs CI(S) per molecular
dynamics time step as a function of the number of particle/hole
states.  In both cases, the effort to construct the two body
interactions is included; however, the CI(S) does {\em not} include
the effort required to compute energy gradients.
As the number of particle/hole states
increases, this becomes the dominant part of the effort. }
\label{fig:cputimes}
\end{figure}

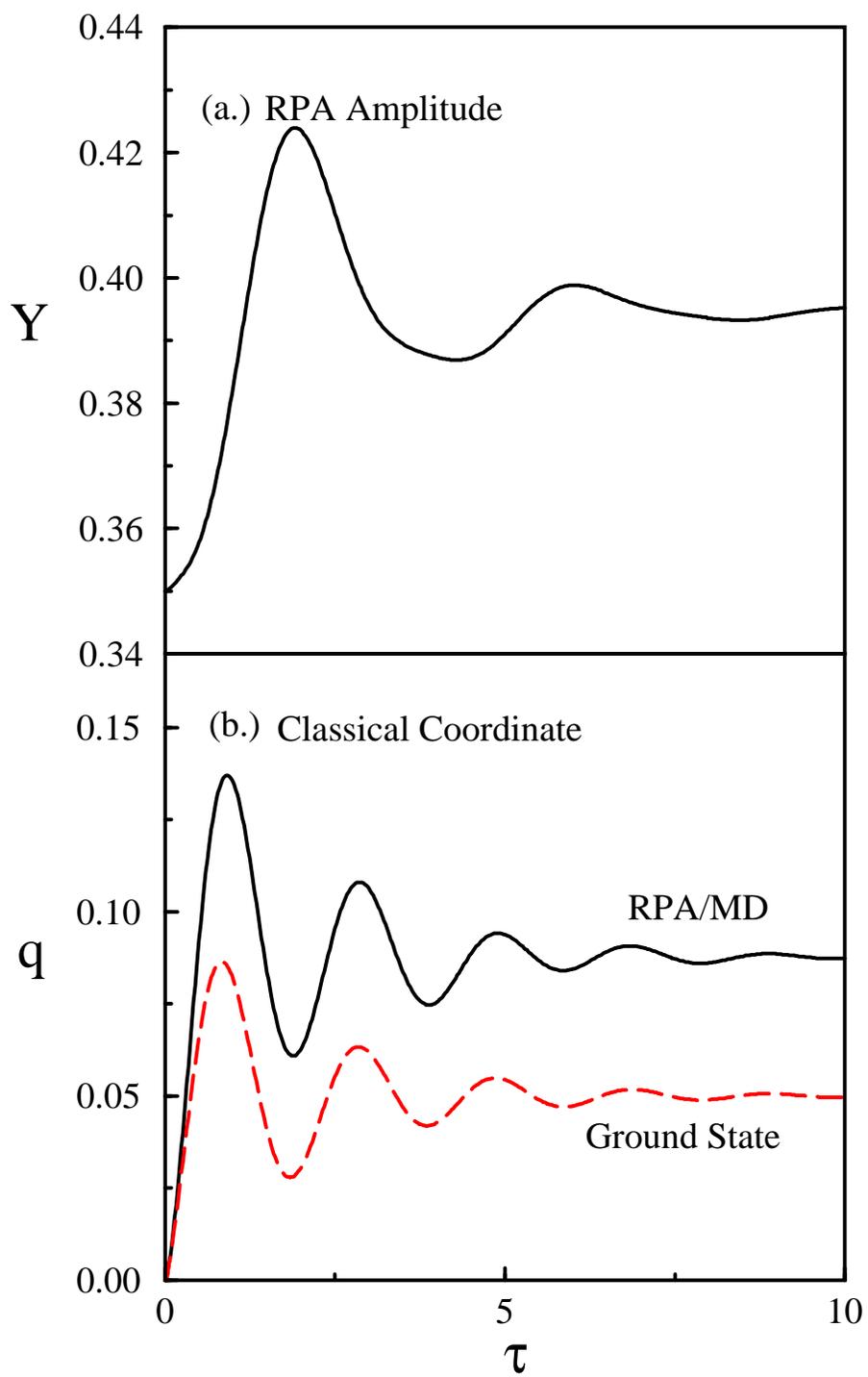
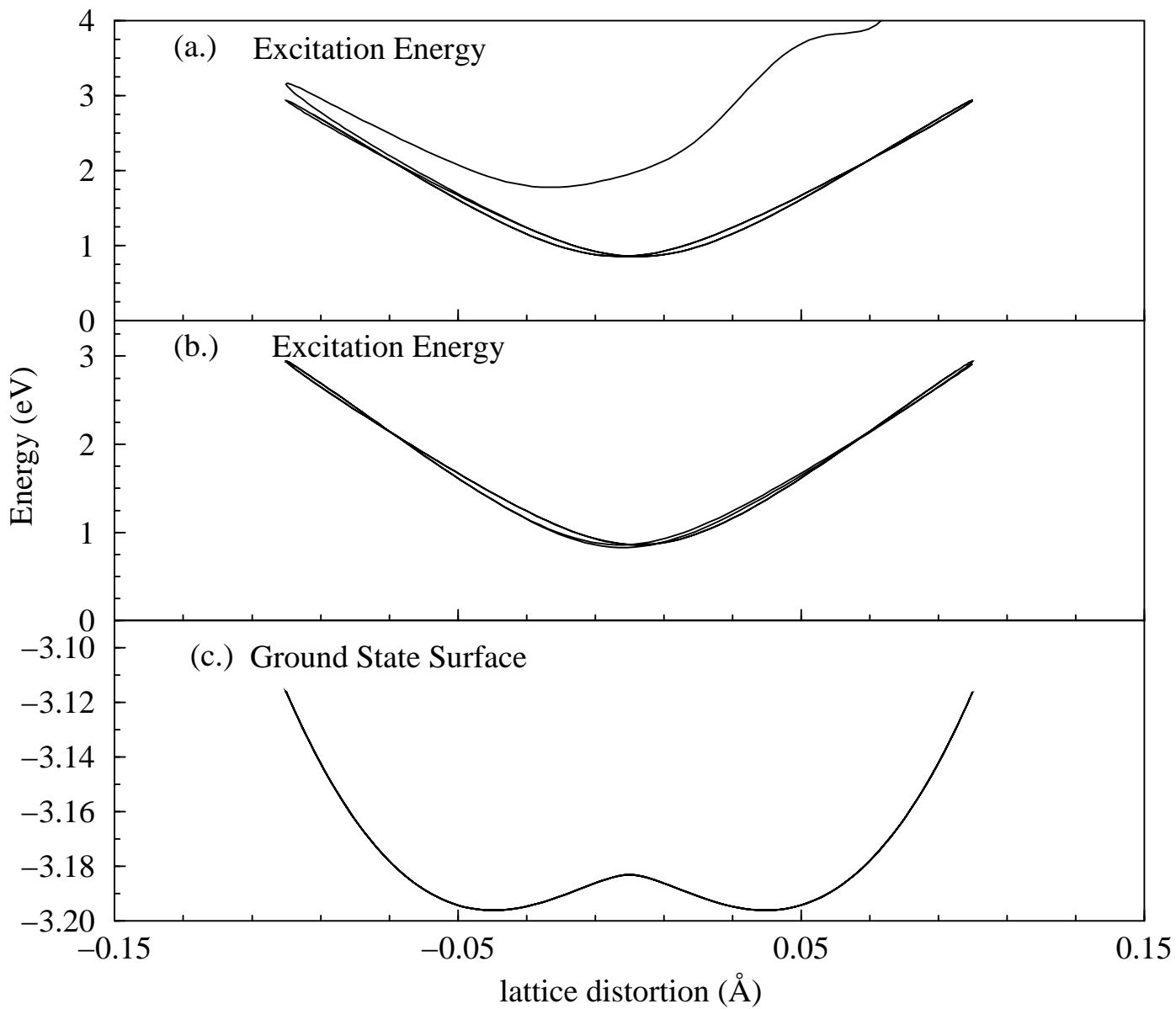
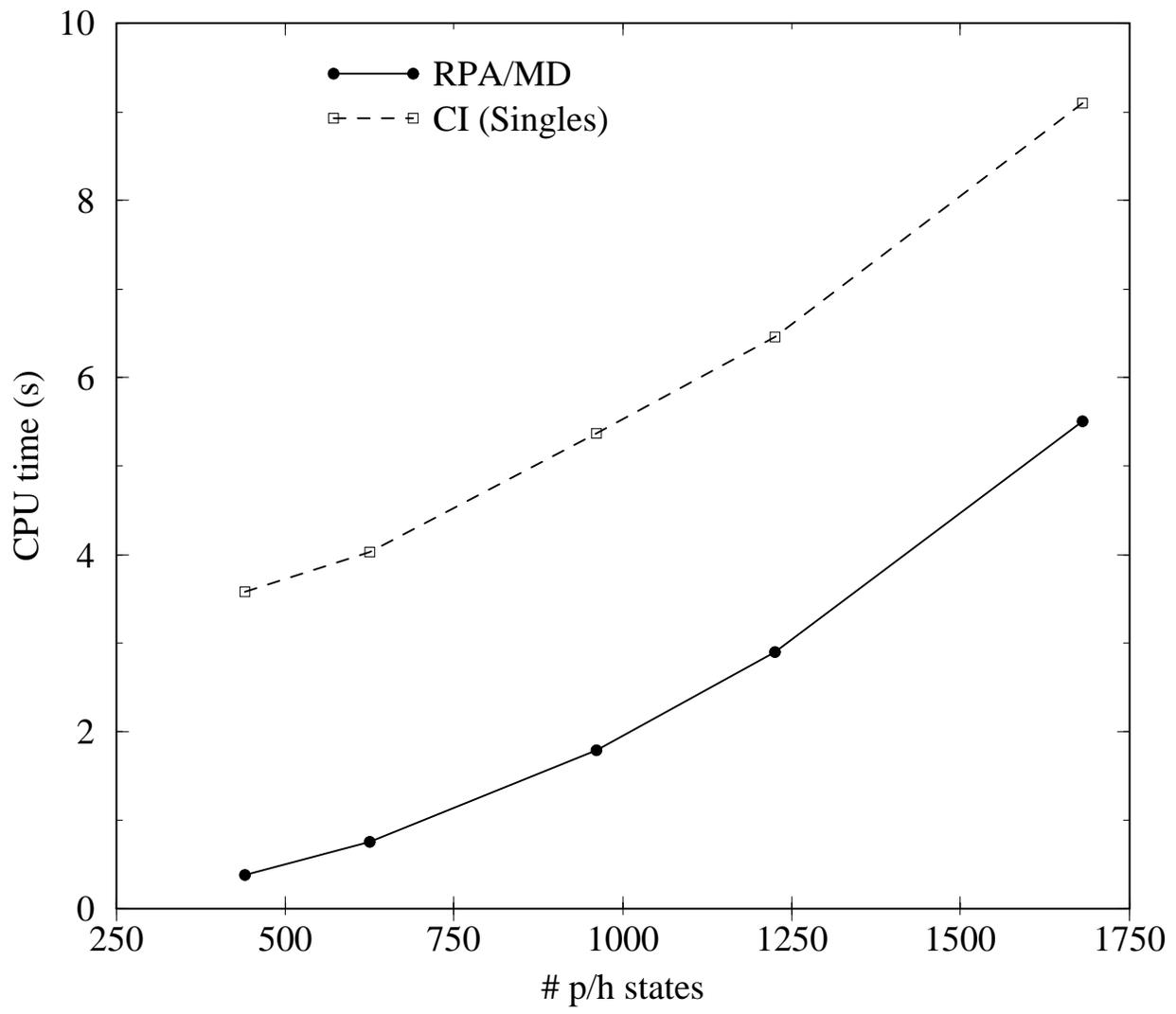
\end{document}